# Distributed Temperature Sensing: Review of Technology and Applications

Abhisek Ukil, *Senior Member, IEEE*, Hubert Braendle, and Peter Krippner

*Abstract*—Distributed temperature sensors (DTS) measure temperatures by means of optical fibers. Those optoelectronic devices provide a continuous profile of the temperature distribution along the cable. Initiated in the 1980s, DTS systems have undergone significant improvements in the technology and the application scenario over the last decades. The main measuring principles are based on detecting the back-scattering of light, e.g., detecting via Rayleigh, Raman, and Brillouin principles. The application domains span from traditional applications in the distributed temperature or strain sensing in the cables, to the latest "smart grid" initiative in the power systems, etc. In this paper, we present comparative reviews of the different DTS technologies, different applications, standard, and upcoming, different manufacturers.

*Index Terms*—Ampacity, anti-Stokes, Brillouin, cable temperature, distributed strain, DTS, fiber optic temperature sensing, fire detection, leakage detection, locomotive temperature, Raman scattering, sag, Stokes, structural health, transformer hot spot.

## I. Introduction

DISTRIBUTED temperature sensors (DTS) measure temperatures by means of optical fibers. Those optoelectronic devices provide a continuous profile of the temperature distribution along the fiber cable.

DTS systems were conceptualized in the 1980s, with significant improvements in the technology and the application scenario over the last decades. DTS systems are capable of measuring temperature with a high degree of accuracy over significant distances. Typical figures are detection accuracy in the range of $\pm 1°C$ at a resolution of $0.01°C$ to a spatial resolution of 1 m, over measurement distances as long as 30 km. Being insensitive to the electromagnetic interference (EMI), the fiber optic-based DTS systems are of particular interest in the electrical applications, e.g., power systems, cables, etc.

In this paper, we present a comparative review of the different DTS technologies and different applications, standard and upcoming, relative to the time of writing. Quasi-continuous methods like a multitude of fiber bragg gratings inscribed in a fiber are not considered in this review. The remainder of the paper is organized as follows. Section II provides overviews about the different technologies of DTS. Section III describes applications in the temperature and strain detection in cables. Various power systems applications are discussed in Section IV. Section V reviews different upcoming applications. Major manufacturers of the DTS systems are mentioned in Section VI. Section VII discusses about the current and future trends in DTS, followed by conclusions in Section VIII.

## II. DTS Technology

DTS technology was invented more than 20 years back. The main measuring principles are based on detecting the back-scattering of light, e.g., using the Rayleigh [1], Raman [2], and Brillouin [3] principles. This section provides a brief overview of the different technologies.

### A. Raman Scattering

Optical fibers are typically made from doped quartz glass, a form of silicon dioxide ($SiO_2$). With the amorphous solid structure of $SiO_2$, the thermal effects along the fiber cause lattice oscillations. As light falls on the thermally excited molecular oscillations, the photons of the light particles and the electrons of the molecule undergo an interaction, resulting in scattered light, also known as Raman scattering. The scattered light has a spectral shift equivalent to the resonance frequency of the lattice oscillation [2], [4].

The back-scattered light contains three spectral components, the Rayleigh scattering with wavelength of laser source, the Stokes component with the higher wavelength in which the photons are generated, and the Anti-Stokes components with a lower wavelength. The intensity of the Anti-Stokes band is temperature dependent, while the Stokes band is temperature insensitive. The ratio of the Anti-Stokes and the Stokes light intensities provides the local temperature measurement [5]. This is shown in Fig. 1. The approach was developed in the 1980s at Southampton University, U.K. [2].

### B. Brillouin Scattering

Brillouin scattering refers to the scattering of a light wave by an acoustic wave due to a nonelastic interaction with the acoustic phonos of the medium [3], [6]. The Brillouin scattering produces both frequency down- (Stokes) and up-shifted (Anti-Stokes) light, given by

$$\nu_b = \frac{\omega_b}{2\pi} = \frac{2nv_a}{\lambda_L} \qquad (1)$$

where, $\nu_b$ is the Brillouin frequency shift, $\omega_b$ is the angular frequency shift, $n$ is the refractive index of the fiber, $v_a$ is the longitudinal acoustic velocity for the fiber, and $\lambda_L$ is the free-space wavelength of the pump light [7]. The Brillouin frequency shift

Manuscript received March 22, 2011; revised June 22, 2011; accepted July 08, 2011. This work was supported by the Sensors & Signal Processing (SSP) program, ABB Corporate Research.

A. Ukil and H. Braendle are with ABB Corporate Research, Baden Daettwil CH-5405, Switzerland (e-mail: abhisek.ukil@ch.abb.com; hubert.braendle@ch.abb.com).

P. Krippner is with ABB Corporate Research, Ladenburg, Germany (e-mail: peter.krippner@de.abb.com).

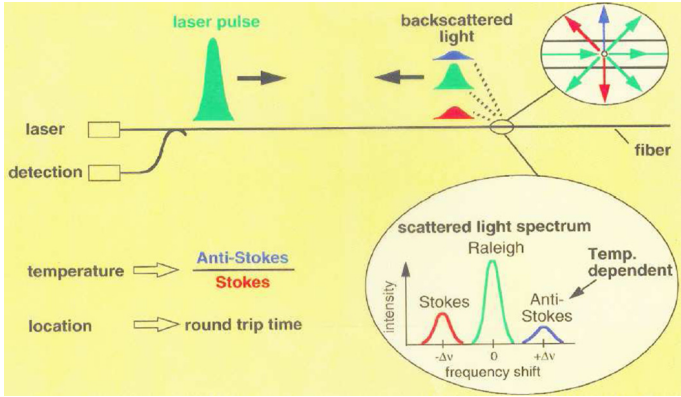

Fig. 1. DTS using Raman scattering derived by the ratio of the Anti-Stokes to Stokes band intensities.

TABLE I
COMPARISON OF DIFFERENT DTS TECHNOLOGIES

| Scattering | Rayleigh [1] | Raman [2] | Brillouin [7] |
|---|---|---|---|
| Temp. sensitivity (%/°C) | 0.54 | 0.8 | 0.01 |
| Temp. range (°C) | 5 to 110 | 0 to 70 | -30 to 60 |
| Accuracy (°C) | 1 | 10 | 1 [6] |
| Spatial resolution (m) | 1 | 3 | 3-5 [6],[10] |
| Fiber length range (m) | 170 | 1000 | 51000 [6] |
| Measurement time (s) | 2.5 | 40 | 4 |
| Strain ($\mu$m) | - | - | 100 [37] |

varies linearly with the strain and the temperature, as given by [7]

$$\partial \nu_b = C_{\nu\varepsilon} \partial \varepsilon_z + C_{\nu\theta} \partial \theta. \qquad (2)$$

For $\lambda_L = 1.55$ $\mu$m, the strain and the temperature coefficients of the Brillouin frequency shift, due to the strain $\varepsilon_z$ and the temperature $\theta$, are measured as ($C_{\nu\varepsilon} = 0.048$ MHz/$\mu$m) and ($C_{\nu\theta} = 1.1$ MHz/K), respectively [8]. Thus, the Brillouin scattering-based technique can be used for sensing both distributed temperature and strain, but not simultaneously both. The different coefficients would be utilized to separate the effects, provided one is interested to measure either the temperature or the strain, but not in a mixed-mode. Even though Parker et al. [7] studied possible ways to simultaneously detect temperature and strain using the Brillouin scattering, that is not quite standard yet.

The Brillouin scattering-based approach was developed mainly in the 1990s. The scattering effect can be seen in the time-domain, e.g., by Brillouin Optical-fiber Time Domain Analysis (BOTDA) [9]. Otherwise, the scattering effect can be done in the frequency domain, e.g., by Brillouin Optical-fiber Frequency Domain Analysis (BOFDA) [10]. The BOFDA usually results in higher spatial resolution [11].

### C. Other Fiber Optic Methods

Sang et al. [12] used Rayleigh back-scattering to measure distributed temperature in a nuclear reactor, using 2 m segments of commercially available single mode optical fibers and frequency domain reflectometry. The Rayleigh scattering-based method claims to measure with an accuracy of 0.6% full scale with a spatial resolution of 1 cm up to 850°C [12].

Chaube et al. [13] used BOTDA technique for dynamic strain measurement. However, traditional BOTDA technique requires time-durations in the range of minutes to make measurements. Their approach aims to reduce the measurement time by new signal configuration using multiple pump frequencies in the form of a frequency-domain comb. They reported measuring the strain distribution of a 120 m long SMF-28 fiber cable in 256 $\mu$s, or at a rate of 3.9 kHz [13].

Spammer et al. [14] used a Sagnac interferometer merged with a Michelson interferometer to detect distributed strain or temperature along the fiber cables. The two interferometers are illuminated by two light sources. The output of the Michelson interferometer is proportional to the phase change caused by the perturbation, while the Sagnac interferometer would be proportional to the product of the phase change by the perturbation and the distance to it. The distance of the perturbation (e.g., temperature or strain profile) could thus be obtained by dividing the output of the Sagnac interferometer by the Michelson one [14].

### D. Use of Rare Earth Ions

Loss of the fiber at a wavelength on the edge of an absorption band can be monitored by measuring the local fiber absorption using backscattering techniques. Rare-earth materials like holmium, erbium, ytterbium, and praesodymium show high sensitivity for utilization in DTS systems. Yataghene et al. [15] reported high thermal sensitivity of holmium-doped optical fibers. Holmium ions show best thermal sensitivity in cryogenic temperature range, around 650 nm [15].

Ko et al. [16] reported DTS systems using the temperature dependent gain of 1.48 $\mu$m pumped erbium-doped fiber amplifier. The reported temperature sensitivity of the gain is $-0.21\%/°C$ for a gain of 28 dB at 20°C. The sensor configuration is similar to the optical time-domain reflectometry, for a potentially cheaper DTS configuration [16].

### E. Comparison

Table I shows a comparative summary of the different DTS technologies developed more or less chronologically in time, namely, the Rayleigh, Raman, and Brillouin scattering. From Table I, even though the Rayleigh scattering shows the best accuracy, it is quite limited in terms of range of fiber length which is very important these days to monitor longer lengths of cables, e.g., in underground power distribution systems. From that point of view, the Brillouin scattering provides best length range, with highest temperature sensitivity and relatively good measurement time. Brillouin scattering could also detect distributed strain, which cannot be done by the other two methods. However, one could not possibly measure the distributed temperature and strain simultaneously. Typically, the applications of the Brillouin scattering are either for distributed temperature measurement or strain, not both simultaneously. Therefore, Brillouin scattering would likely be a preferred choice in future developments, as a replacement for the Raman scattering as already seen in the commercial systems [11].

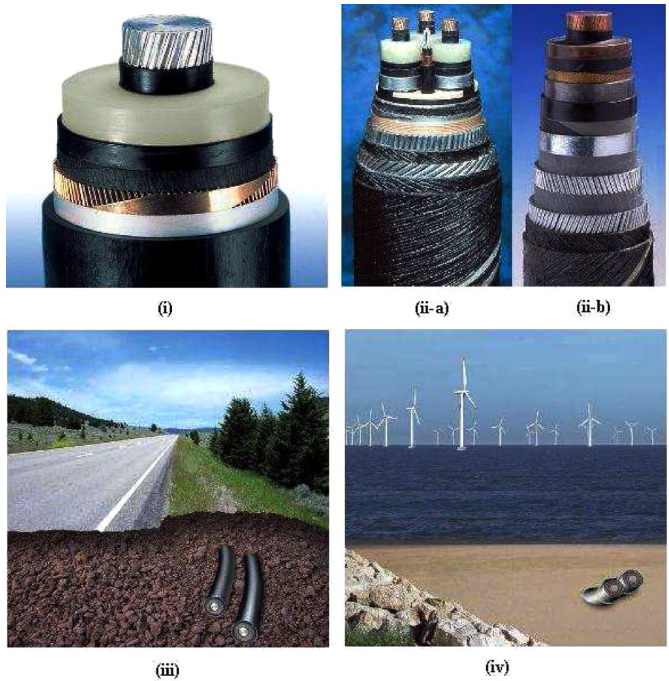

Fig. 2. Power cables: (i) XLPE (420 kV), (ii-a) submarine XLPE (420 kV), (ii-b) submarine mass-impregnated (600 kV), (iii) HVDC Light (320 kV), and (iv) HVDC Light submarine (320 kV).

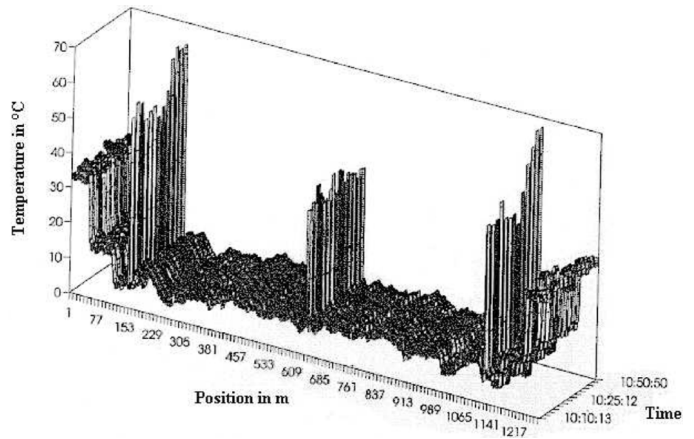

Fig. 3. Temperature profiling of underground cables, showing temperature distribution along the cable in meter resolution taken at different time.

## III. Cable Applications

### A. Power Cables and Utilizations

High- and medium-voltage (HV and MV) cables are important parts of the power transmission and distribution systems. Starting from as early as 1883, until date cables for different voltage levels are available, e.g., from 36 kV (MV) to extra high voltage (EHV) levels like 420 kV or 500 kV for HVDC applications [17]. Power cables are increasingly becoming important for more and more underground power distribution systems as they are considered safe, robust, and optimized for space compared to overhead lines [18]. Fig. 2 shows different classes of HV cables [17].

### B. Cable Temperature and Ampacity

The maximum current (Amp) carrying capacity of an underground cable transmission or distribution system is limited by the maximum allowable conductor and cable surface temperature. This is termed as ampacity [19]. The surrounding soil composition, density, and moisture content have influence on the heat dissipation capability of the cable. Thus, even if the cable has good quality and thermal performance, the inherent inhomogeneous ambient conditions cause thermal unbalance and hot spots along underground cables. Thus, continuous thermal profiling along the length of the cable is of particular interest. And DTS is inherently applicable in this context.

Nakamura *et al.* [20] reported temperature monitoring of 275 kV EHV $1 \times 2500$ mm$^2$ aluminium-sheathed XLPE cable joints in underground power transmission lines. They made comparative surface temperature measurements using DTS (Raman scattering based) and several thermocouples along the cable. Real-time measurements using the DTS system showed significant accuracy within the acquisition time frame of 1 min.

A Raman scattering-based DTS system was applied by Kawai *et al.* [21] on 77 kV 400 mm$^2$ XLPE cables to detect fault points along the cable at an alarm threshold of 5°C. The optical fiber was placed in a stainless steel sheathed tubular structure of the cable.

ABB studies on 150 kV cable transmission system were reported in [22], using Raman scattering-based DTS. Example measurement, shown in Fig. 3, demonstrates the temperature distribution along the cable with a resolution of 1 m taken at different time. The accuracy of the measured temperature was in the range of $\pm 1$°C, with measurement time of 5–10 s depending on the distance [23]. The spatial positions between 533 to 609 m show temperature hot spots in the cable. This type of temperature hot spots limits the overall ampacity of the cable. However, accurate temperature profiling can quantitatively evaluate the limits of the ampacity which is often set empirically without knowing it. Thus the cable's ampacity does not get utilized to its full potential [24]. Typical layout configurations of fiber integrated cables are shown in [25].

### C. Other Cable Related Applications

Tayama *et al.* [26] used DTS systems to detect anchor damage and defacement of wire armor in a 6.6 kV XLPE submarine cable laid in the seabed between Kata and Tomogashima in Japan. Different optical fibers were used for monitoring the cable temperature (fiber put in the cable core) and the mechanical damage (fiber put in the outer bedding of core).

The optimal usage of DTS systems for cables depend on the conductor temperature as well as prediction of the maximum allowable ampacity towards safe overloading specification. In the direction towards the latter, Li [27] used DTS systems to estimate the soil thermal parameters from the measurement of the surface temperatures of the cables. From the DTS measurements, soil parameters like thermal diffusivity, conductivity, etc. are estimated using finite-element methods and the gradient-based optimization method.

## IV. Power Systems Applications

As already mentioned in the previous section, long range DTS was developed essentially due to the need of monitoring long undersea cables, requiring spatial resolution around 10 m [28]. This section reviews other power systems applications.

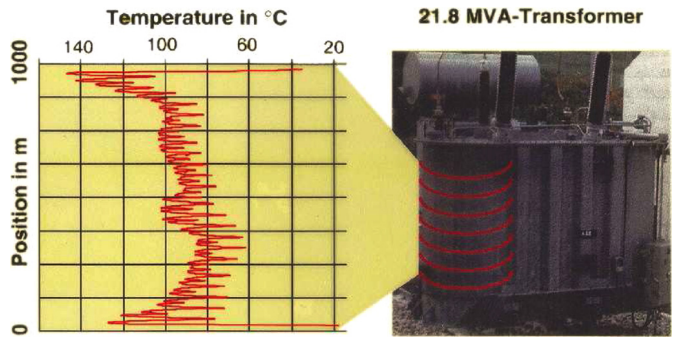

Fig. 4. Fiber optic distributed temperature measurements along the windings in power transformer.

### A. Transformer Monitoring

Transformers are one of the most important as well as expensive equipments in the power systems. Not being a rotating element, cooling of the transformer windings is very important, which is often done via oil, air (fan), or combined ways. However, nonuniform localized temperature rise can cause rapid thermal degradation of insulation. Analytical methods of estimating the temperature distribution and prediction of the hot spots are discussed in [29] and [30]. Nevertheless, DTS systems are quite relevant in this aspect to provide a real temperature distribution along the transformer winding. Currently, DTS is a tool for validation of thermal models during transformer development and for use in type tests rather than a broad usage of online monitoring system.

Studies [22], [25], [31] report successful tests using DTS systems (Raman scattering based) to measure the temperature of a 22 MVA, oil-cooled power transformer. The fiber was housed in a v-grove (mm range) with paper isolation in the copper wire which was used in the winding, the total run length being in the range of 1000 m. Thus, the DTS was able to measure the exact temperature profile in the transformer winding. The temperature distribution along the winding of the 22 MVA transformer is shown in Fig. 4. From Fig. 4, the variations of the temperature can be noticed, especially the higher temperatures at the both ends of the winding. This was due to the fact that at the end part of the winding, the inhomogeneity of the magnetic field causes increased eddy current losses. The temperature fluctuations noticeable throughout the winding shows the nonuniform reach of the cooling oil to the inner and the outer depths due to physical layout of the winding. Nevertheless, the detailed temperature profile provides a precise condition monitoring of the transformer winding.

### B. Traction Transformer Monitoring

Monitoring of the temperature in a traction transformer was also reported in [22] and [31]. For this, DTS system was used to monitor the temperature variations of the winding in a traction transformer in a locomotive (Lok 2000, type 460, no. 107) in Switzerland. Temperature was monitored in the transformer winding as the 607 ton locomotive traveled in a route Zurich-Gotthard-Chiasso in Switzerland, with variations in altitudes. This gave different loading conditions of the transformer. The measurements are shown in Fig. 5. In Fig. 5, the mid-part between Zurich and Chiasso has hilly regions which yielded in higher temperatures.

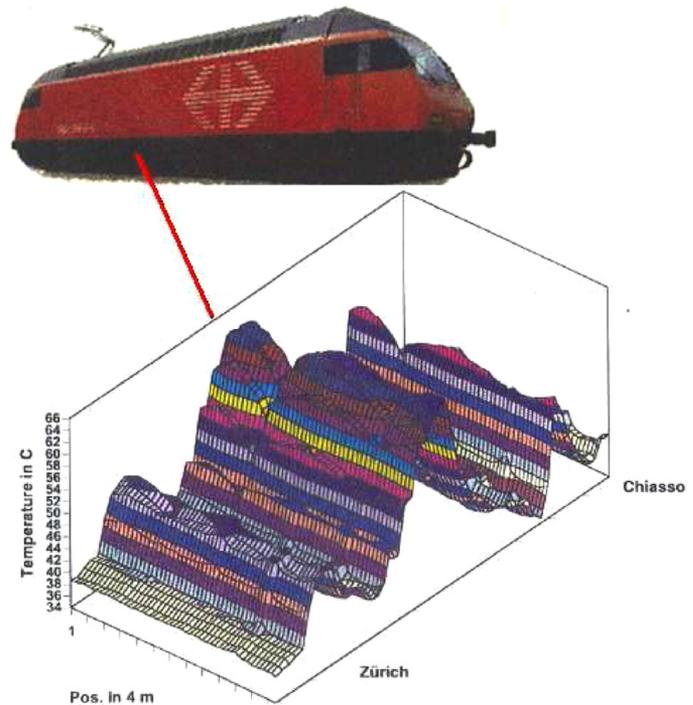

Fig. 5. Temperature measurements along the windings in a traction transformer in locomotive.

### C. Switchgear Monitoring

Boiarski *et al.* [32] reported monitoring of distributed temperature in other power plant devices, e.g., switchgear at R.E. Burger plant, West Virginia. Commercially available DTS system was installed in a 2.4 kV switchgear cell. Starting at 14.7 m, the fiber was attached to the lower finger cluster in the switchgear with a 2.1 m fiber loop between the two cluster regions [32]. The DTS system used Rayleigh backscattering. From the temperature distribution at an accuracy of $\pm 5°C$, it could be noticed that the maximum temperature in the switchgear occurs near the top part of the bottom cluster [32].

### D. Rotating Machine Monitoring

A DTS system was used to monitor the rotor winding temperature of a synchronous condenser at Pacific Gas and Electric substation in San Mateo, California [32]. Four fiber optic sensors, epoxied to prevent vibration effects, were inserted in the separate V-block regions of the rotor. The distributed temperature measurements were in agreement with the simulated data [32].

DTS system-based monitoring of rotor winding temperature in a boiler feed-pump motor was investigated at the Hudson Avenue power station, Con Edison Company, New York [32]. The DTS system successfully monitored the winding temperature at different loading conditions of the motor. It also measured stator and air passage temperature rise above ambient as a function of the motor current [32].

### E. Pipework Monitoring in Generating Plants

In Pressurized Fluidized Bed Combustion (PFBC) power generating plants, hot output gases ($\sim 850°C$) from different reactors are brought together, passing through a ceramic filter, to drive the turbines for power generation [28]. The temperature of the corrosive gases flowing under high pressure significantly exceed the operating limit of the pipes which are insulated internally. Therefore, it becomes critically important to monitor the temperature of the pipeline to detect early any problem in the insulation, welded joints, seals, etc. DTS systems were successfully applied in a fiber loop of approximately 4000 m to monitor the temperature. The specially designed high-temperature fiber optic sensor was laid along the steel surfaces. This would be similar to about 3000 point temperature measurements [28].

*F. Surface Monitoring of Reformer Vessel*

Reformer vessel is a brick-lined steel vessel in which a conversion process is carried out at high temperature (∼2000°C) and pressure. If there are any failures in the internal burner, the flame could get disoriented, causing serious damage to the refractory bricks. A DTS system was applied to monitor the entire surface of the vessel for hot spots. A total of about 1700 m fiber loop was used to successfully monitor the temperature of the surface of the vessel, indicating temperature variations in the range of ±50°C across different parts of the vessel [28].

*G. Monitoring of Boiler Furnace*

Lee *et al.* [33] reported the development of a DTS system for real time monitoring of high temperature in boiler furnace in power plants. The interesting parameters are the spatial and the temporal distributions of high temperatures within a gas, coal, or oil-fired boiler. These play an important role in assessing and controlling the polluting sources like NOx.

*H. Monitoring of Overhead Transmission Lines*

Overhead lines are an indispensable part of the power transmission and distribution (T&D) systems. For modernizing the aging T&D systems, the utility industry is putting a lot of focus on reliable, flexible, and efficient systems in the "smart grid" initiative. The ampacity standard of the overhead lines are one of the most important factors that influence such initiative.

In T&D system, the ampacity of the overhead lines is governed by the current versus temperature characteristics [34]. Therefore, it becomes highly important to estimate the temperature distribution along the lines, which would be also influenced by weather and environmental factors like ambient temperature, wind speed, any snow condition, etc. Nevertheless, the proper understanding of the temperature profiles along the overhead lines would enable the operators to extend the current carrying capacity within the safety limit. The temperature distribution in the overhead conductors would also influence the sag and tension of the conductors.

The work by Klein *et al.* [35] reports development of a reliable software-based sag, tension, and ampacity measurement system, called "STAMP". The program provides conductor temperature, sag, and tension in real time, taking into account weather conditions like wind speed, wind direction, air temperature, etc. The software-based approach is aimed at providing lowcost solution. In their work, Klein *et al.* made use of single-mode DTS system as a benchmark to calibrate and compare the accuracy of the software program. They reported 95% confidence levels in estimating the temperature, sag, and tension, vis-á-vis the DTS-based measurements [35].

As mentioned above, the amount of power flow in a conductor is limited by the ampacity of the same. Excessive high temperature might cause faster aging as well as cause the conductors to elongate, exceeding the sag safety limits above the ground. Thus, sag monitoring is a very important aspect, typically performed by measuring the surface temperature of the phase conductor, measuring conductor tension, measuring using image-processing software, using global positioning systems (GPS), measuring the angle of the conductor at the pole, etc. [36].

Utilizing the sag measurements, Olsen and Edwards [36] proposed a lowcost method to indirectly estimate the average conductor core temperature. The principle is based on attaching two ends of a grounded wire of high electrical resistance to an appropriate location on each of the two transmission-line towers, then measuring the induced current on the wire by the transmission-line conductors [36].

The average conductor core temperature $\theta_c$ could be estimated by the following relation [36]:

$$\theta_c = \theta_r + \frac{(s - s_r)}{\alpha}\left[\frac{s + s_r}{6L} - \frac{wL^2}{2ss_r ae}\right] \quad (3)$$

where $\theta_r$ is the reference temperature for which the reference sag $s_r$ is known, $s$ is the measured sag, $w, a, e, \alpha$ are the conductor's weight per unit length, cross-sectional area, effective modulus of elasticity, and temperature coefficient of linear expansion, respectively, and $2L$ is the length of the tower span [36].

Overall, in the overhead line monitoring applications, the DTS systems have to compete against model-based solutions which use only a few temperature measurement points, line sag measurement, etc. Here as well, currently DTS is more a tool during development rather than for online monitoring, applications like load balancing, etc. for cost reasons.

## V. OTHER APPLICATIONS

*A. Structure Monitoring*

Lanticq *et al.* [37] reported design and validation experiments to monitor strain and temperature in concrete structures using fiber optic cables based on Brillouin scattering. They used both standard single-mode fiber and polarization maintaining fiber embedded into a reinforced concrete beam, and the Brillouin time-domain method (BOTDA). The acceptable strain measurement results show promise towards continuous health monitoring in structural environments.

This is more and more needed to prevent accidents like the sudden collapse of I-35W Mississippi river bridge in August 2007 [38]. Continuous health monitoring sensors laid along the concrete structures could indicate incipient damages, difficult to pick up by human inspection, leading to catastrophic failures [39].

*B. Leakage Detection*

In civil engineering, health monitoring of dams and dikes is very important to identify early any leakage in order to prevent catastrophic failures like flooding, etc. The conventional methods are based on manual inspection which is rather unreliable. Studies by Khan *et al.* [40], [41] report the utilization of DTS systems to monitor leakage.

The principle relies on the fact that due to leakage, significant water flow causes temperature change between the canal water and the ground. The DTS system then effectively provides the distributed temperature profile along the dam.

However, the change in temperature could also be caused by other factors like seasonal variation, precipitations, etc. Therefore, source separation analysis [40] is also required to reliably use the thermometric data from the DTS.

Nikles *et al.* [42] used Brillouin scattering-based DTS systems to perform leakage detection in a 55 km brine pipeline in Berlin, Germany. The reported accuracy is 1°C over the 55 km with a measuring time in less than 10 min [42].

Sensornet case-studies [51] report utilization of DTS systems to detect leakage in gas and water pipelines for the Korean National Gas authority, monitoring of a 57 km Ethylene pipeline in northern Germany, monitoring any leakage in a liquefied natural gas (LNG) pipeline in an oil and gas company in Texas, etc.

### C. Application in Oil and Gas

DTS systems are also applied in the oil and gas industry. In oil/gas well, DTS installations are typically categorized as retrievable, semipermanent, and permanent. Details about such installations can be referred to in [43].

DTS is used to monitor the temperature log of the well. The temperature log could then be associated with the effects of the liquid flow when a well is shut in. Gas entry into a channel causes a cooling effect which can be detected in the static rathole. The DTS temperature log could also provide information on the cooling effect in a water injection well. DTS systems are also used to monitor wells on a pad or production platform, steam breakthrough in a production well. DTS systems could also be used to detect the start up of electric submersible pumps (ESP) and motor [43].

### D. Fire Detection

Automatic fire detection is an important topic to prevent asset damage and human casualty. Automatic fire detection systems are based on optical smoke detector, ionization detector, infrared or ultraviolet flame detection, etc. One of the most promising technologies is linear optical fire detection, based on the DTS principle, e.g., Raman scattering [44]. Using frequency domain reflectometry in a silica glass fiber, the study [44] reported tracking of rapid changes in the temperature profile to detect fire and localize the seat of the fire within a building with 1 m resolution.

### E. Applications in Mines

For safe electrical operation, in mines, shuttle car trailing cables should be operated below the safety limit which is about 90°C [45]. Operation in temperature condition more than that might cause premature insulation failure. Therefore, temperature monitoring under dynamic condition is very important.

Dubaniewicz *et al.* [45] reported utilization of DTS systems embedded within the metallic conductors to measure temperatures at 1 m intervals along the entire cable length, at an accuracy of ±1°C.

In addition to the distributed temperature, Dubaniewicz *et al.* [46] also reported use of fiber optic technology to monitor atmospheric related parameters like methane ($CH_4$) and carbon monoxide (CO).

## VI. Manufacturers

The list of manufacturers of DTS systems is a growing one. Some of the leading manufacturers are mentioned below. However, this is not an exhaustive list, and the order does not signify any relative measure.

- Sensa® [47]: DTS systems like SUT-® family, DTS-® family for applications like power T&D, leakage detection in oil and gas, fire detection, etc.
- omnisens® [48]: distributed monitoring systems for temperature, strain, fatigue, etc., like DITEST-® LTM®, AIM®, SHM®, DSM®, STA-R®, DLIGHT series®, etc.
- es&s® [49]: DTS system like DiTeSt® for distributed temperature and strain measurement.
- LIOS TECHNOLOGY® [50]: stand alone DTS systems, integrated Real Time Thermal Rating (RTTR) package.
- sensornet® [51]: fiber optic sensors and digital monitoring systems.
- SensorTran® [52]: DTS product families like ASTRA®, CENTUARUS®, GEMINI®, NEPTUNE®, CABLES®.
- Weatherford® [53]: optical DTS, optical pass-thru pressure/temperature gauge.
- AP SENSING® [54]: DTS system like EN54-5®.
- Promore Engineering Inc.® [55]: reservoir monitoring systems.
- sabeus® [56]: Field Sense™ MPT temperature sensing systems, BHPt pressure sensing systems.
- LUNA Technologies® [57]: Distributed Sensing System™ (DSS) 4300 for making distributed measurements of temperature and strain.
- HALLIBURTON® [58].
- MAXIM® [59].

Besides the manufacturer companies, there are some nonprofit professional societies as well.

- The Fiber Optic Association Inc. [60].
- Subsea Fiber Optic Monitoring Group [61].
- IEEE Photonics Society [62].

## VII. Discussions and Future Trends

The following comments are cited.
1) The cost of DTS systems and installations vary greatly for different applications and settings. For example, in the power systems and cable applications often the voltage level, type of cable, distance, etc. are the influencing factors. Typical costs could be in the range of USD 100 000. In oil and gas applications, also the cost depends highly on the well configurations, distance, applications, etc., typically being in the range of USD 50 000–150 000 [43]. Therefore, the cost of the DTS systems is still on the higher side.
2) Therefore, one obvious future need would be to have cost-effective DTS solutions maybe in the range of USD 10 000–20 000, or less.
3) Despite the relatively higher cost, the real-value of DTS systems is to provide distributed temperature and/or strain measurement. This is often useful for safety critical or difficult to access applications, involving equipments which themselves are highly expensive, e.g., HV power transformer, subsea oil and gas applications, etc. Getting a direct and continuous measurement of the temperature profile to prevent potential damage to such equipments is a growing trend in continuous condition monitoring for predictive maintenance.
4) Therefore, a better way of judging the value of DTS-based monitoring would be to introduce alternative business models based on total cost of ownership (TCO) for expensive devices. TCO would incorporate device costs, life cycle and operating costs, costs including all additional infrastructure, maintenance, replacement in case of no monitoring, etc.
5) With the "smart grid" initiative, integration of renewables and distributed energy sources like off-shore wind farms, remote photovoltaics, etc. are foreseen. This would imply increased utilizations of HVDC cables, etc. Synonymously, usage of the DTS systems for monitoring the cable health, optimal usage of the cable current carrying capacity, better loading of the cables, overhead conductors for energy efficient operations would be increasingly important.
6) Over the last decades, the DTS interface softwares have evolved greatly with flexible graphical user interfaces, faster acquisition time, interoperability amongst the different vendor equipments, etc.
7) With the advancement of Brillouin scattering-based DTS systems, structural health monitoring of concrete structures, buildings, dams, etc. are possible. This is very important to prevent catastrophic failures [38]. Also, continuous monitoring of leakages along water, oil, and gas pipelines over several kms would be an advantage to improve the system or plant efficiency and reliability.
8) The fiber optic-based DTS systems are inherently insensitive to electromagnetic interference, etc. This is of particular interest in the electrical applications, e.g., power systems, cables, etc.

## VIII. Conclusion

DTS systems measure temperatures by means of optical fibers, detecting via Rayleigh, Raman, and Brillouin principles. Using Brillouin scattering, it is also possible to measure the distributed strain. DTS systems provide a continuous profile of the temperature that is distributed along the cable or other device. In this paper, we presented a comparative view of the different standard DTS technologies along with nonstandard ones. On application side, we reviewed distributed temperature sensing in cables for better ampacity judgment. In the power systems, monitoring of power transformer and traction transformer in order to detect hot spots is very important. Monitoring of the power plant devices like switchgear, rotating machine windings, boiler furnace, reformer vessel, etc. are reviewed as well. Leveraging advancement of Brillouin scattering, applications like structural health monitoring in concrete structures, leakage monitoring in dams, oil and gas application, fire detection, and safety applications like in mines are reviewed. The number of companies providing DTS systems is growing ever so much, some being listed herein. Overall, the DTS systems provide great advantages in terms of continuous profiling of distributed temperature and/or strain for safety critical or difficult to access applications.

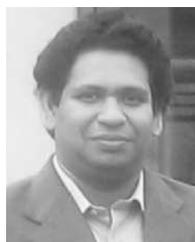

**Abhisek Ukil** (S'05–M'06–SM'10) received the bachelor of electrical engineering degree from the Jadavpur University, Calcutta, India, in 2000, the M.Sc. degree in electronic systems and engineering management from the University of Bolton, Bolton, U.K., in 2004, and the Ph.D. degree from the Tshwane University of Technology, Pretoria, South Africa, in 2006, working on power systems disturbance analysis with Eskom.

From 2000 to 2002, he worked as a software engineer at InterraIT, India. After joining in 2006, currently he is a Principal Scientist at the Integrated Sensor Systems Group, ABB Corporate Research, Baden-Daettwil, Switzerland. He is author/coauthor of more than 40 refereed papers, a monograph *Intelligent Systems and Signal Processing in Power Engineering* (New York: Springer, 2007), two book chapters, and inventor/co-inventor of six patents. His research interests include signal processing, machine learning, power systems, embedded systems, sensors, and sensor-based systems.

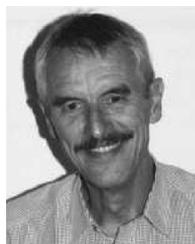

**Hubert Braendle** received the Ph.D. degree in physics from the University of Zurich, Zurich, Switzerland, in 1974.

From 1974 to 1978, he was a Postdoctoral Researcher in high-energy physics with the University of California, Los Angeles, and with the Swiss Federal Institute of Technology Zurich, Zurich, Switzerland. In 1978, he joined BBC (now ABB Switzerland, Ltd.), Baden, Switzerland. He is currently a Research Fellow at the Sensor Technologies Group, ABB Corporate Research Center, Baden-Daettwil, Switzerland, where he is engaged in research on sensor technologies, optical sensing, and analytics.

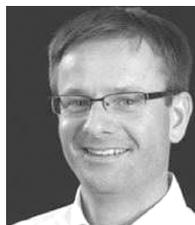

**Peter Krippner** received the Ing. degree in electrical engineering from the Karlsruhe Institute of Technology, Karlsruhe, Germany, in 1995, and the Ph.D. degree in mechanical engineering from the Karlsruhe Institute of Technology, Karlsruhe, Germany, in 2000.

He was a group leader at the Karlsruhe Institute of Technology dealing with micro optical sensing devices. He joined ABB Corporate Research, Ladenburg, Germany, in 2000, working as senior principal scientist in the area of field instruments and gas and liquids analyzers with a focus on micro electro mechanical systems. Since 2007, he has been heading the global ABB R&D program for sensing and signal processing, covering the whole range of sensing technologies and signal processing techniques to create innovative products as well diagnostic functionality in the area of condition monitoring.